\documentclass[10pt,letterpaper]{article}
\usepackage[top=0.85in,left=1.6in,footskip=0.75in]{geometry}

\usepackage{amsmath,amssymb}

\usepackage{changepage}

\usepackage[utf8x]{inputenc}

\usepackage{textcomp,marvosym}

\usepackage{cite}

\usepackage{nameref,hyperref}

\usepackage[right]{lineno}

\usepackage{microtype}
\DisableLigatures[f]{encoding = *, family = * }

\usepackage[table]{xcolor}

\usepackage{array}

\usepackage{caption,tabularx,booktabs}

\newcolumntype{+}{!{\vrule width 2pt}}

\newlength\savedwidth

\usepackage[aboveskip=1pt,labelfont=bf,font=small,labelsep=period,singlelinecheck=off]{caption}

\makeatletter
\renewcommand{\@biblabel}[1]{\quad#1.}
\makeatother

\usepackage{lastpage,fancyhdr,graphicx}
\usepackage{epstopdf}
\begin{document}
\vspace*{0.2in}

\begin{flushleft}
{\Large
\textbf\newline{\textbf{Temporal connection signatures of human brain networks after stroke}}
}
\newline
\\
Catalina Obando\textsuperscript{1,2,*},
Charlotte Rosso\textsuperscript{2,3,4},
Joshua Siegel\textsuperscript{5},
Maurizio Corbetta\textsuperscript{6,7}
Fabrizio De Vico Fallani\textsuperscript{1,2,*}
\\
\bigskip
\textbf{1} Inria Paris, Aramis Project Team, Paris, France
\\
\textbf{2} Institut du Cerveau et de la Moelle epiniere, ICM, Inserm U 1127, CNRS UMR 7225, Sorbonne Universite, Paris, France
\\
\textbf{3} AP-HP, Urgences Cerebro-Vasculaires, Hopital Pitie-Salpetriere, Paris, France
\\
\textbf{4} ICM Infrastructure Stroke Network, STAR team, Hopital Pitie-Salpetriere, Paris, France
\\
\textbf{5} Department of Psychiatry, Wasingthon University, St Louis, US
\\
\textbf{6} Department of Neuroscience and Padova Neuroscience Center, University of Padova, Italy
\\
\textbf{7} Department of Neurology, Wasingthon University School of Medicine, St Louis, US
\\
\bigskip

%
%
%



* corresponing authors: catalina.obando@gmail.com ; fabrizio.devicofallani@gmail.com

\end{flushleft}
\bigskip

\section*{Abstract}

\begin{small}
Plasticity after stroke is a complex phenomenon initiated by the functional reorganization of the brain, especially in the perilesional tissue. 
At macroscales, the reestablishment of segregation within the affected hemisphere and interhemispheric integration has been extensively documented in the reconfiguration of brain networks and proved to be a potential biomarker of functional recovery. 
However, the local connection mechanisms generating such global network changes are still largely unknown as well as their potential to better predict the outcome of patients. To address this question, time must be considered as a formal variable of the problem and not just a simple repeated observation. 
Here, we hypothesize that the temporal formation of basic connection blocks such as intermodule links -or edges- and intramodule connected triads -or triangles- would be sufficient to determine the large-scale brain reorganization after stroke.

To test our hypothesis, we adopted a statistical approach based on temporal exponential random graph models (tERGMs). First, we validated the overall performance on synthetic time-varying networks simulating the reconfiguration process after stroke. 
Then, using longitudinal functional connectivity measurements of resting-state brain activity, we showed that both the formation of triangles within the affected hemisphere and interhemispheric links are sufficient to reproduce the longitudinal brain network changes from 2 weeks to 1 year after the stroke. 
Finally, we showed that these temporal connection mechanisms are over-expressed in the subacute phase as compared to healthy controls and predicted the chronic language and visual outcome respectively in patients with subcortical and cortical lesions, whereas static approaches failed to do so.
Our results indicate the importance of considering time-varying connection properties when modeling dynamic complex systems and provide fresh insights into the network mechanisms of stroke recovery.
\end{small}



\newpage

\section*{Introduction}


The brain is a networked system whose parts dynamically interact over
multiple temporal and spatial scales. Such network properties are at the
basis of neuroplasticity allowing for the acquisition of new skills
(e.g., learning) as well as for the functional recovery after brain
injuries (e.g., stroke). When locally damaged, the brain tends to
spontaneously adapt by recruiting new resources through the network to
compensate for the loss of the neuronal tissue and recover the
associated motor or cognitive functions. At micro/meso spatial scales,
dendritic remodeling, axonal sprouting, and synapse formation have been
best demonstrated in the peri-lesional tissue of animal models during
the first weeks after stroke~\cite{Mostany_2011,Brown_2008,Thomas_Carmichael_2001}.~ At larger macro scales, it has been shown that post-stroke plasticity also involves regions outside
the peri-infarct cortex - including the contralesional hemisphere - and
that the associated brain activity and connectivity changes can last
several months in an effort to return to a normal
condition~\cite{Carrera_2014,Siegel_2016,Weiller_1992,Dancause_2005,van_Meer_2010,He_2007}.

Recovery after stroke is a temporally dynamic network phenomenon, but
only recent longitudinal studies have allowed to demonstrate a direct
association between changes in brain functional connectivity (FC) networks and
spontaneous recovery in humans~\cite{Ramsey_2016}. Both increased
interhemispheric homotopic integration and intrahemispheric segregation
appear to be fundamental principles for recovery of associative/higher
cognitive functions (e.g. attention, language), as quantified by the
return to a normal modular organization~\cite{Siegel_2018}. Indeed, the
role of time in complex networks has been recently revised as a
fundamental variable to model and analyze real-world connection
phenomena~\cite{Li_2017,Holme_2012}. 
By introducing time, new higher-order properties emerge that cannot be captured by static network, or graph, approaches. In a temporal network two disconnected nodes can, for example, still interact if there exists a time-ordered sequence of links connecting their neighbors \cite{Holme_2012,Li_2017}. This way of rethinking networks naturally extend standard topological properties with purely temporal concepts such as latency, persistence, or \textit{formation} of connectivity patterns or motifs.
Interestingly, the inherent ability of temporal-topological graph metrics to characterize dynamic brain networks, as well as to predict future behavior, have been increasingly demonstrated in
the context of human neuroscience~\cite{Bassett_2011,Tang_2010,Thompson_2017,Braun_2016,Cole_2013,Ekman_2012}.~

To date, however, the existence of dynamic brain network signatures in
stroke and their ability to predict functional outcome in individual
patients has not been proved directly. Based on these theoretical and
empirical grounds, we hypothesized that temporal connection mechanisms
constitute fundamental building blocks of the recovery process after
stroke. More specifically, we expected that the formation of interhemispheric links and
intrahemispheric clustering connections, and not just their static appearance, would
characterize the crucial phases of brain reorganization after stroke. Furthermore, we
hypothesized that these temporal connection signatures in the subacute phase would allow to predict the future behavior of patients in the chronic phase.

To test this hypothesis, we considered longitudinal brain networks
derived from resting-state fMRI BOLD activity recorded in a group of
patients at 2 weeks, 3 months and 1 year after their first-ever
unilateral stroke. Neurological impairments were described using
multidomain behavioral measurements at each visit. We evaluated the
significance of the hypothesized temporal connection mechanisms through
a rigorous statistical network approach based on temporal exponential random graph models (tERGMs). We compared our results with respect to a group of demographically-matched healthy subjects and we tested their ability to predict the future outcome of stroke patients.
See \textbf{Material and Methods} for more details on the experimental
design and methods of analysis.

\section*{Results}
\subsection*{A statistical model of temporal networks}

It has been theoretically hypothesized and recently empirically observed
that recovery after stroke is accompanied by the emergence of FC patterns supporting interhemispheric integration and intrahemispheric segregation
(\textbf{Fig 1A}). To quantify these dynamic network properties we introduced
two temporal graph metrics which account respectively for the formation
of edges between hemispheres ($E$) and triangles -i.e., interconnected triads- within hemispheres
($T$) (\textbf{Material and methods}). Both $E$ and $T$ quantify connection patterns through time that cannot be derived by looking within single time steps (\textbf{Fig 1B}). Without loss of generality, time steps corresponded here to different longitudinal visits, or sessions.

\begin{figure}[h!]
\begin{center}
\includegraphics[width=1.00\columnwidth]{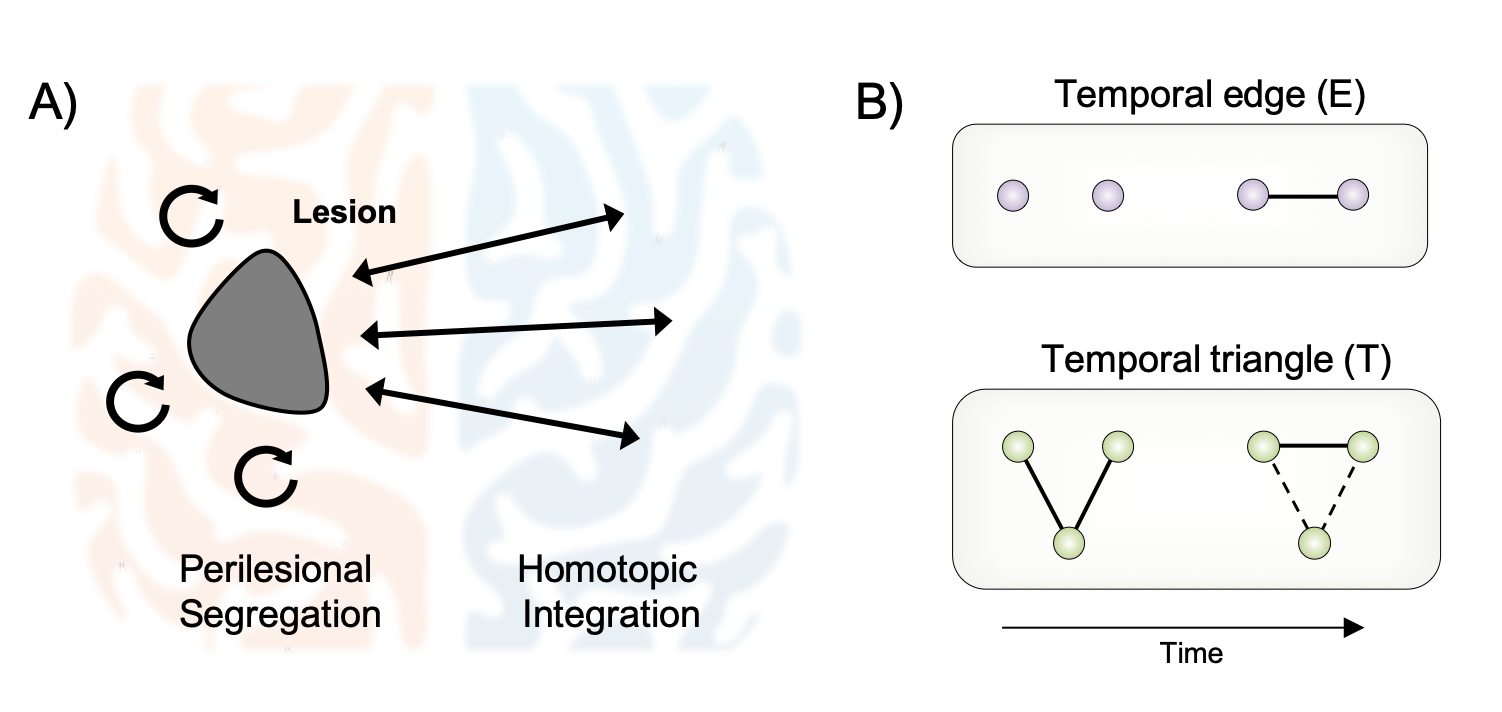}
\caption{{\bf Dynamic network hypothesis of brain reorganization after stroke}
A) Graphical representation of the known resting-state functional connectivity (FC) changes occurring after stroke. Brain networks tend to restablish functional segregation within the affected hemisphere (orange) as well as functional intergation with the homotopic areas in the unaffected hemisphere (blue) \cite{Siegel_2018}. 
B) Temporal graph metrics to quantify the hypothesized local mechanisms underlying the global network FC changes, i.e. formation of interhemispheric links (temporal edge $E$) and clustering connections within the affected hemisphere (temporal triangle $T$). Both $E$ and $T$ quantify connection patterns taking place across consecutive time steps. Hence, the closure of a triangle from time $t-1$ to $t$, does not necessarily imply the existence of a complete interconnected triad as illustrated by the dashed edges in $T$.
}
\label{fig1}
\end{center}
\end{figure}

To validate the relevance of these metrics we first generated synthetic
dynamic graphs that reproduce the known topological changes observed in resting state brain networks after stroke~\cite{Siegel_2016,Siegel_2018}. The model consisted of a
network of two modules - representing the hemispheres - that were
initially not interconnected and randomly initiated with a fixed connection density. This condition
ideally reproduce the effects of stroke with complete destruction of
between-hemisphere integration and within-hemisphere segregation. To
simulate the reorganization process underlying recovery, a fraction of links was randomly rewired in each subsequent time step to form inter-module edges ($E$) and intra-module
triangles ($T$) (\textbf{Fig 2A}). To simulate realistic time
dependencies~\cite{Holme_2012,Li_2017}, we applied the rewiring rule in a way
that each network at time~\(t\) could be obtained by the
network at time~\(t-1\) (\textbf{Materials and methods}).

By construction, the cumulative occurrence of temporal edges $E$ and
triangles $T$ in the synthetic networks increased over time (\textbf{Fig
2B}). To evaluate the extent to which these connection mechanisms were
also sufficient for statistically reproducing the observed networks we
adopted a general framework based on temporal exponential graph models
(tERGM) ~\cite{Hanneke_2010}. The probability of observing the entire
series of networks was here controlled by two main
parameters,~\(\theta_E\) and~\(\theta_T\), which weighted
the relative contribution of~\(E\) and~\(T\). Because tERGMs are in practice fitted through numerical approximations, we also introduced two secondary parameters to ensure the convergence to an optimal solution, namely the connection density and stability (\textbf{Material and methods}).

Results showed a high goodness-of-fit in terms of link prediction capacity $\left\langle AUP \right\rangle =0.8(0.02)$ and $\left\langle AUR \right\rangle=0.88(0.01)$ (\textbf{Fig 2C}), which we did not obtain when using static graph metrics merely counting the appearance of edges and triangles $\left\langle AUP \right\rangle=0.22(0.01)$ and $\left\langle AUR \right\rangle=0.59(0.05)$. Global dynamic network changes were determined by the local formation of temporal edges and triangles as indicated by the positive values of the corresponding parameters meaning that they occur more often than expected by chance ($\theta_T=0.13$, $\theta_E=0.16$).
Furthermore, we showed that the networks sampled with the fitted tERGM
also captured the global integration and segregation changes, here quantified
by a decreasing network modularity $Q$~\cite{Newman_2006}, which were not directly included in the model but indirectly resulting from the
simulated process. Instead, the use of static metrics which simply account for the presence of edges and triangles in each time step failed to retrieve the actual trend (\textbf{Fig 2D}, \textbf{Supp. text}) .

\begin{figure}[h!]
\begin{center}
\includegraphics[width=1.00\columnwidth]{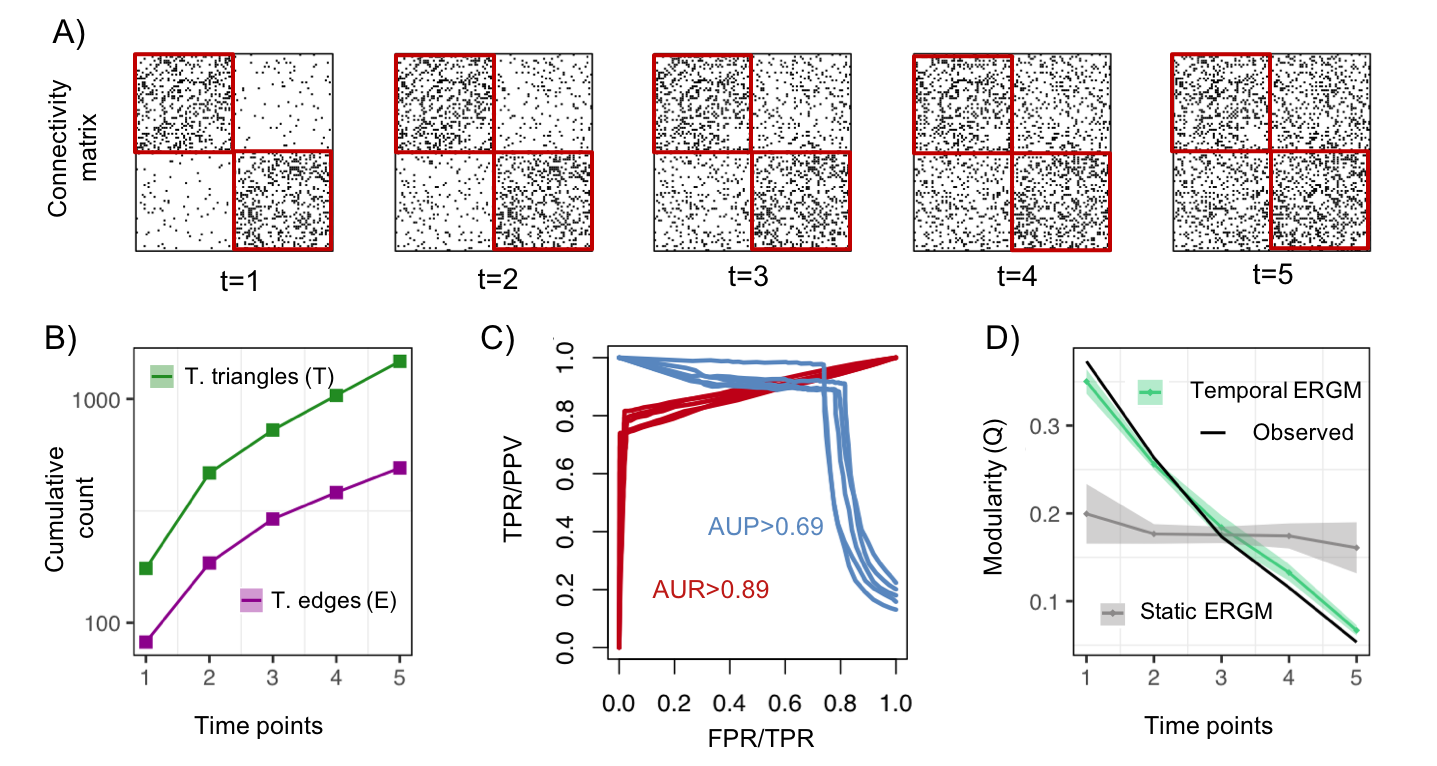}
\caption{{\textbf{Validation of temporal exponential graph models (tERGMs) on synthetic networks.} 
A) Sequence of connectivity matrices representing a time-varying synthetic network. Unweighted and undirected networks are generated by a model that reproduces the large-scale reconfiguration changes after stroke. Interhemispheric integration and within-hemisphere segregation are obtained by allowing the formation of temporal edges $E$ between blocks and temporal triangles $T$ within blocks. Blocks (i.e. red squares) represent the hemispheres (\textbf{Material and methods}).
B) Cumulative counts of $E$ and $T$ between consecutive time steps for the synthetic temporal network. 
C) Prediction performance: area under the curve for the receiver operating characteristic (AUR, red curve) and precision recall (AUP, blue curve) of the out-of-sample link prediction in the networks simulated by the tERGM. Different curves correspond to different time steps.
D) Validation: modularity $Q$ values for \textit{i)} the actual time-varying synthetic network (black line), \textit{ii)} the ones simulated by the tERGM (green line), and \textit{iii)} those simulated by a static version of the tERGM where the graph metrics were only accounting for the presence - not the formation - of edges and triangles (gray line) (\textbf{Material and methods})
{\label{fig2}}%
}}
\end{center}
\end{figure}

\par\null

\subsection*{Dynamic connection mechanisms after stroke}

{\label{dynamicMech}}

We next considered longitudinal brain networks underlying functional
recovery in unilateral stroke patients (\textbf{Material and methods}). Brain
networks were obtained by computing functional connectivity between fMRI
activity of different regions of interest (ROIs) (\textbf{Fig
3A})\cite{Gordon_2014}. Damaged ROIs and the corresponding connections were excluded by the network analysis (\textbf{Materials and methods}). Stroke lesions involved both subcortical and cortical ROIs, with cortical lesions mainly covering
cingulo-opercular, auditory, ventral-attention and default mode network
systems (\textbf{Fig 3B}). While the number of occurrences was slightly higher for the left hemisphere, the average size of the lesion was larger for the right hemisphere, i.e. $5,6$ cm$^3$ \textit{versus} $4,7$ cm$^3$ (\href{Table_Demo_clin.xlsx}{\textbf{File S1}}). %

Previous analysis demonstrated a progressive restoration of modularity in the large-scale functional brain network after stroke that was associated with good recovery (\textbf{Fig 3C,D})\cite{Siegel_2018}. 
These network changes were more evident in the sub-acute phases (from 2 weeks to 3 months), but, critically, they were obtained separately in each session thus ignoring the time-ordered nature of the reorganization process.
Here, we used tERGMs to statistically identify the temporal reconfiguration mechanisms after stroke and evaluate their ability to predict the future behavioral outcome of patients. 
In this experimental setting, the temporal graph metrics $E$ and $T$ now quantify respectively homotopic integration between the hemispheres and segregation within the lesioned hemisphere. 
To increase the specificity of our approach, we restricted the analysis to the subnetwork corresponding to the affected system for patients with cortical lesions ($n=23$), while we considered the whole hemisphere as affected for subcortical-lesioned patients ($n=26$) (\textbf{Materials and methods}). 

\begin{figure}[h!]
\begin{center}
\includegraphics[width=1.00\columnwidth]{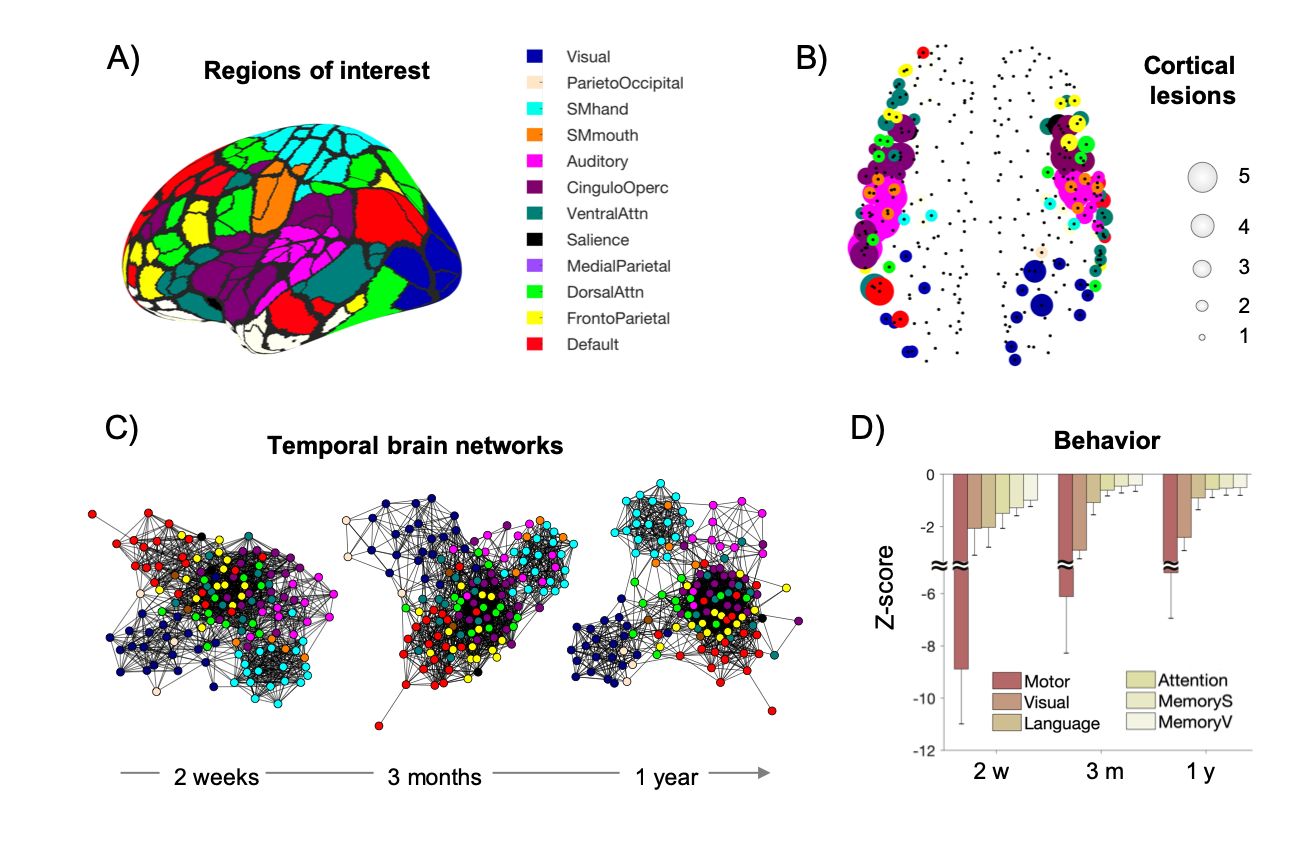}
\caption{{\textbf{Longitudinal brain networks and behavior after stroke. } 
A) Cortical parcellation used to define the nodes (i.e. ROIs) of brain networks \cite{Gordon_2014}. Different colors highlight the belonging to different functional systems.
B) Spatial position of the ROIs affected by the stroke and their occurrences in the cortical-lesioned subgroup of patients ($n=23$). $55,82\%$ of the damages are in the left hemisphere; $44,18\%$ are in the right hemisphere. The average size of the lesion is however larger for the right hemisphere ($5,6$ cm$^3$ \textit{versus} $4,7$ cm$^3$).
C) Dynamic brain network of a stroke patient over time. Nodes are spatially arranged according to a spring layout. Network are filtered and binarized to retain the $10\%$ of the links with strongest functional connectivity (FC) values.
D) Longitudinal multidomain behavioral scores for the stroke population including both cortical- and subcortical-lesioned patients. Only patients with score in every visit are reported here ($n=23$). Values are Z-normalized with respect to a demographically-matched control group ($n=21$). Bars show the group-averaged values and whiskers indicate standard~error means (SEM). Behavior is significantly recovering in all domains but visual and regardless of the damage location, i.e. cortical/subcortical (Repeated-measures ANOVAs, $p<0.05$, \href{Table_ANOVA.xlsx}{\textbf{File S3}}). 
{\label{fig3}}%
}}
\end{center}
\end{figure}

After fitting a tERGM to each patient, we found a general high goodness-of-fit regardless of the observation window (i.e. 3 months or 1 year) and location of the lesion (i.e. subcortical/cortical)(\textbf{Tab S1}). 
Sometimes, the parameter coefficients were not assigned because the respective counts were not sufficiently represented in the network. This mainly occurred for the \(\theta_T\) coefficients of cortical-lesioned patients, whose network were relatively small  (\textbf{Materials and methods}). These situations were excluded from any subsequent analysis. 
We then focused on the subacute phase (from 2 weeks to 3 months), where most changes were previously observed \cite{Siegel_2018}. The main parameter values \(\theta_E\) and \(\theta_T\) were in average positive indicating that both the formation of interhemispheric connections and triangles within the lesioned hemisphere are peculiar mechanisms of brain network reconfiguration after stroke (\href{Table_ERGM_coeff.xlsx}{\textbf{File S2}}).
We validated this result in the subcortical-lesioned group by showing that both $\theta_E$ and $\theta_T$ values were significantly higher compared to those obtained in a group of demographically-matched healthy controls 
(Wilcoxon test $p<0.01$, \textbf{Fig 4A}, \href{Table_Demo_clin.xlsx}{\textbf{File S1}}). 

A similar group comparison could not be performed for cortical-lesioned patients because tERGMs were fitted to the subnetwork affected by the lesion, and there was a high variability in the lesion locations and size across patients (\textbf{Material and methods}). Nevertheless, such variability allowed us to elucidate the relationship between the observed dynamic network mechanisms and the cortical system that was affected. 
Results showed that different systems were not responding in the same way to the lesion and that primary cortical areas reacted with a higher formation of temporal edges and triangles. 
Specifically, we found higher \(\theta_E\) values when the stroke involved visual and sensorimotor systems (\textbf{Fig 4B}). \(\theta_T\) values were higher in visual and default-mode systems prevalently in the right hemisphere, whose damages were in average larger compared to the left hemisphere. However, we reported no significant relationships between the size of the lesion and the tERGM coefficients.

Taken together, these findings unveil the specific local connection mechanisms giving rise to global network segregation within the lesioned hemisphere and interhemispheric homotopic integration after stroke.

\subsection*{Prediction of future outcome}

{\label{PredOutcome}}

Given the dynamic nature of brain connectivity after stroke, we finally asked whether such temporal network properties could predict the future outcome of patients. 
To do so, we correlated the values of the tERGM parameters fitted over the dynamic brain networks in the
subacute phase with the multi-domain behavioral scores gathered in the chronic phase (1 year)  (\textbf{Materials and methods}).
Due to patient dropouts, some behavioral scores were missing. 
For the correlations analysis we only considered patients with all scores in all the visits (\textbf{Fig 3D}).

For subcortical lesions ($n=14$), we found that both the formation of interhemispheric temporal edges ($\theta_E$) and intrahemispheric temporal triangles ($\theta_T$) significantly predicted the future language score (Spearman correlation $R>0.62, p<0.02$, \textbf{Fig 4C}). This tendency was also confirmed in patients with stronger deficit ($n=13$), i.e. when the behavioral score at $2$ weeks was at least one SD below the mean of the healthy group ($R=0.74, p=0.005$ for $\theta_E$; $R=0.64, p=0.021$ for $\theta_T$). 
For cortical lesions ($n=9$), we only reported a significant correlation between $\theta_E$ values and the visual score in the chronic phase ($R=0.73$, $p=0.031$), which however we could not assess for the few patients with greater impairment ($n=3$).
No other significant predictions were found for the other behavioral scores (\textbf{Tab S2}). 

In a separate analysis, we verified that these predictions could be obtained neither when we calculated the temporal graph metrics outside the tERGMs nor when we considered equivalent static graph metrics for $E$ and $T$, by neglecting the past networks (\textbf{Supp. text}). 
Notably, language and visual outcomes after 1 year were not predicted by the precedent values at 2 weeks or by the relative difference between 3 months and 2 weeks. In addition, no significant correlations were reported when we considered the lesion size as predictor. Only the age factor exhibited a weak correlation with the future visual outcome in cortical-lesioned patients ($R=0.56, p=0.04$), but it was not a significant predictor for the language outcome in the subcortical-lesioned group. 

These results indicate that the statistical occurrence of temporal connection mechanisms reflecting within-hemisphere segregation and interhemispheric homotopic integration after unilateral stroke, might be also crucial for the prediction of functional recovery.



\begin{figure}[h!]
\begin{center}
\includegraphics[width=1.00\columnwidth]{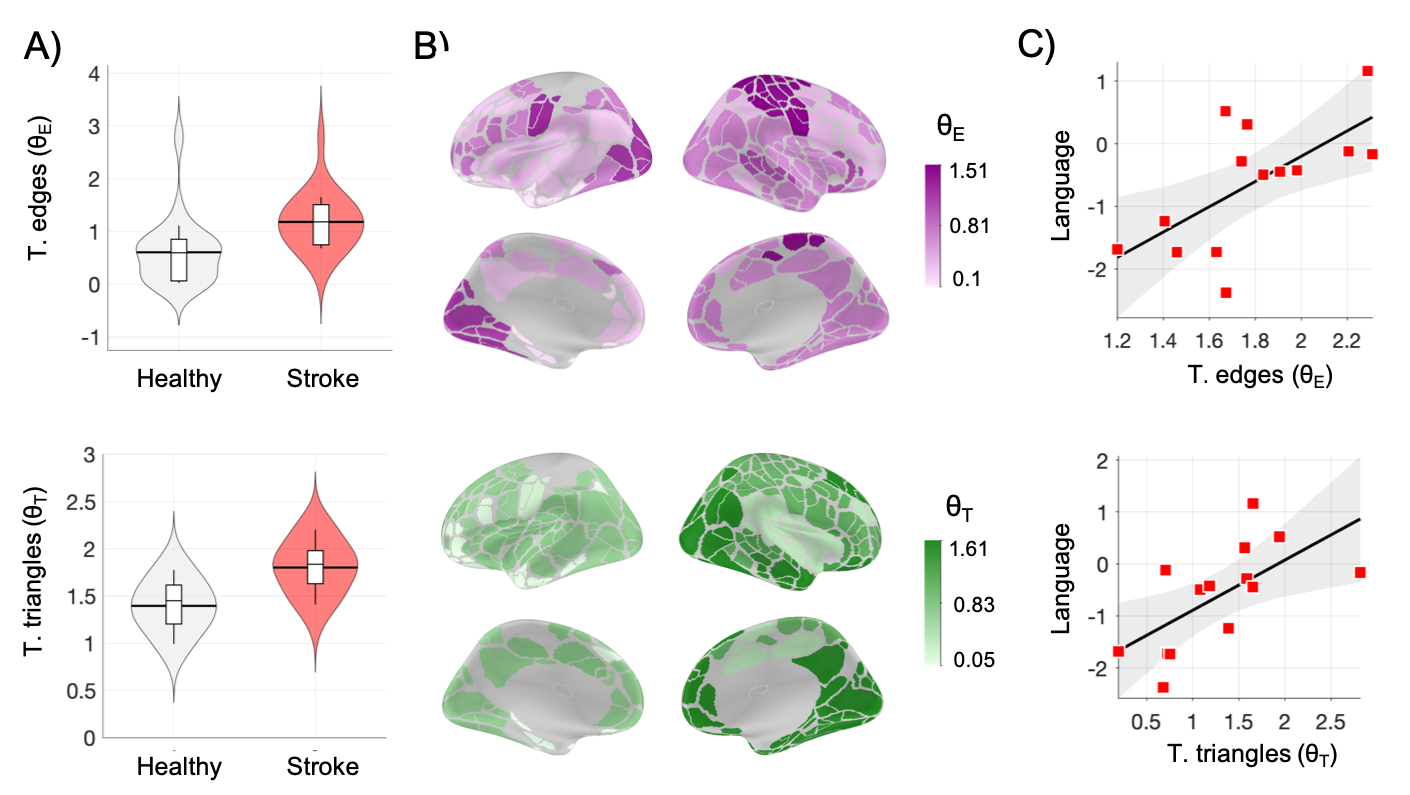}
\caption{{\textbf{Temporal connection mechanisms after stroke and prediction of outcome}
A) Statistical comparison between the main tERGM coefficients of subcortical lesioned patients (red shape) and healthy controls (white shape). Violin plots show the distribution of the values, while innze box-plots denote median and quartiles. Significant increases in the stroke group are reported for both temporal formation of interhemispheric edges ($\theta_E$, Wilcoxon test $t=4.75, p=0.002$) and within-hemisphere triangles ($\theta_T$, Wilcoxon test $t=3.59, p=0.002$).
B) Cortical maps of the main tERGM coefficients for the cortical-lesioned group of patients. Colors denote the average of the parameter values (violet for $\theta_E$, green for $\theta_T$) for the ROIs within the affected functional systems. 
C) Correlation plots of the tERGM coefficients and future outcome in the subcortical-lesioned group. Both temporal edges and triangles in the subacute phase (from 2 weeks to 3 months) significantly predict the global language score in the chronic phase (1 year) normalized to the control group (Spearman correlation $R=0.62$, $p=0.0186$ for $\theta_E$; $R=0.69$, $p=0.006$ for $\theta_T$). 
{\label{fig4}}%
}
}
\end{center}
\end{figure}

\section*{Discussion}

\subsection*{Modeling dynamic networks}

{\label{Discu_modeling}}

Many natural and social interconnected systems are characterized by
time-varying interactions so that the network's structure changes over
time. Such temporality has been shown to affect many dynamical processes
on the network such as slowing down synchronization and diffusion of
information~\cite{Masuda_2013}, impeding exploration and accessibility
\cite{Lentz_2013,Starnini_2012}, as well as favoring system
control~\cite{Li_2017}.

More pertinent to this paper, recent evidence suggests that brain
functional connectivity is inherently dynamic and exhibits relatively fast
fluctuations that support normal cognitive abilities~\cite{Vidaurre_2017,Deco_2017}
as well as slower changes associated to neurodegenerative diseases~
\cite{disease2016} or recovery after brain injuries~\cite{Ovadia_Caro_2013}.
Despite the ubiquity of temporality, brain connectivity networks~have been mostly studied with cross-sectional experiments and static graph approaches. Furthermore, the statistical relevance of the extracted network properties remain largely unknown and group-level analysis is typically used to determine the confidence intervals, thus leading to a critical loss of individual specificity~\cite{De_Vico_Fallani_2014}.

To address these limitations we adopted a model-based statistical framework to test the significance of specific local connection rules to generate an observed sequence of temporal networks. We showed that the temporal formation of local clustering connections and long-range edges, i.e. basic components of segregation and integration of information  \cite{Watts_1998,BOCCALETTI_2006,Sporns_2013}, are sufficient to statistically reproduce longitudinally dynamic brain networks in single stroke patients and reproduce the main global changes.

\subsection*{Brain plasticity and stroke}

{\label{Discu_plasticity}}

Although stroke represents a focal damage, it is well known that
consequences involve areas that are also outside the perilesional
tissue \cite{Mostany_2011,Carrera_2014,Nudo_2013}. Efforts to characterize brain network
reorganization after stroke has focused almost exclusively on the static
representation of underlying connectivity patterns
\cite{Grefkes_2011}. However, both scientific intuition and recent
evidence suggest that temporal network properties might also contain
important information about the mechanisms of brain plasticity
\cite{Bassett_2011}.

Our exploration of temporal network properties provides new insights into the brain organizational principles after unilateral stroke. We found that the formation of clustering connections within the affected hemisphere and of functional interactions with the contralesional hemisphere, constitute fundamental building blocks of cortical plasticity during the initial phases after stroke.
Biologically, these local connection processes can be seen as precursors of the large-scale within-hemisphere segregation and between-hemisphere integration, which have been hypothesized to underlie stroke recovery~\cite{Siegel_2016} and are in line with recent evidence showing a progressive return to a normal modular organization~\cite{Siegel_2018}.~

Specifically, both temporal triangles and edges were evident in the visual system, with temporal edges being also pronounced in the sensorimotor system. These primary systems are known to be more densely connected as compared to other secondary ones~\cite{Narayanan_2005,Reich_2001,So_2011}, with many anatomical fibers crossing the hemispheres for the integration of lower-level visuomotor functions~\cite{Schulte_2010}.
Vulnerability and modeling analyses indicate that attacking such systems will indeed have relatively little effect in terms of widespread connectivity disruption as compared, for example, to midline and fronto-temporal cortices~\cite{Alstott_2009,Honey_2008}. 
This higher structural redundancy would therefore represent a potential reserve of the primary cortical systems to functionally react and reorganize after stroke \cite{Medaglia_2017}.
In parallel, we showed that the higher values of temporal edges and triangles observed in the right hemisphere might not only reflect the lesion size. Patients with right lesions often suffer from severe attention disorders \cite{corbetta_spatial_2011} which cause poorer outcome overall \cite{ween_factors_1996,aszalos_lateralization_2002}, as well as stronger cognitive deficits including language \cite{connor_attentional_2000}. It is possible that the higher propensity of the right hemisphere to reorganize would therefore reflect attentional increases that are known to globally support recovery. 
Further studies will be crucial to elucidate how these temporal connection mechanisms are affected by the lesion side and by the intrinsic FC lateralization \cite{liu_evidence_2009,wang_functional_2014}.

\subsection*{Forecasting behavior and recovery}

{\label{Discu_forecasting}}

Forecasting behavior is paramount in many real-life situations. In clinical neuroscience, a correct prognosis can have concrete impact on the life of people allowing to identify appropriate therapeutics to slow down the progression of disease or promote effective recovery.
Our results show that the intrinsic temporal brain network signatures in the subacute phase after unilateral stroke (from 2 weeks to 3 months) can predict the future outcome of patients in the chronic phase (1 year), whereas static approaches failed to do so.

In the case of subcortical lesions, where the entire hemisphere is concerned, both the temporal formation of interhemispheric links and clustering connections within the affected hemisphere were associated with a better language recovery. These results are in line with previous evidence showing that large-scale functional connectivity changes correlate with cognitive and integrative functions, whereas structural changes such as the lesion location better predict motor deficits~\cite{Siegel_2016,Siegel_2018}. Language disorders can indeed arise not only from the disruption of language processing, but also from the deterioration of distributed support functions including auditory processing, visual attention as in reading, and motor planning for speech \cite{Fedorenko_2014}. 

As for cortical lesions, where specific systems are attacked, the formation of new connections between the perilesional tissue and the homotopic area in the unaffected hemisphere was found to be predictive of the future visual outcome. While it has been demonstrated that lesions to the visuomotor system mainly affect the corresponding functions \cite{Alstott_2009}, the visual pathway involve several remote regions including frontal, temporal, and parietal lobes \cite{Rowe_2013}. Hence, deficits to visual fields can be also given by damages of other areas controlling for example the eye movement or interpreting what we see. This evidence would support the oberved predicition in spite of the heterogenity of the cortical systems that are damaged.

In conclusion, our results highlight the importance of the formation of specific connection mechanisms in the initial phases after stroke. Further research will be crucial to elucidate the intermingled relationship between the damaged brain systems and the affected functions.

\subsection*{Limitations/perspectives}

{\label{Discu_limitations}}

The studied dynamic brain networks consisted of three time points - two
for the subacute phase and one for the chronic phase -~ allowing to have
a partial sampling of the reorganizational mechanisms taking place
after stroke.  
From a methodological perspective tERGMs can fit short network time series and do not make any assumptions as to whether the time that passes between time steps is long or short \cite{Leifeld_2019}. 
Nevertheless, a denser follow-up would have provided more detailed dynamics on the brain network reconfiguration after stroke, but this remains difficult because of the scarce availability of patients to be recruited frequently over long periods.
This is the reason why the large parts of the studies are cross-sectional or only consider two time points \cite{Grefkes_2011,Westlake_2011}. Although the dataset used here is one of the most complete currently
available, longitudinal studies with more frequent visits will be important to assess dynamic
neural mechanisms after stroke at a finer temporal resolution.

The temporal graph metrics implemented in our tERGMs were designed to
capture monotonic network changes over time, such as the formation of
specific connectivity motifs. This means that in general, these models
cannot capture inverse trends - e.g. pattern dissolution - or more
complex dynamics. While this is not a major issue in the case of neural
recovery and neurodegeneration, the study of functional brain networks
at shorter time scales might need more sophisticated approaches to
also model connectivity fluctuations. More research is needed in this
direction and possible solutions may come from the development of
ERGMs with time-varying parameters~\cite{d2017} and
stochastic actor-oriented models~\cite{Kolar_2010}.

The cohort of stroke patients was heterogeneous in terms of stroke
lesion type and location. Because patients suffered from unilateral
lesions - except for brainstem damages - we only considered the
corresponding cortical hemisphere as the affected one. However,
unilateral lesions of subcortical structures including white matter
and cerebellum, might result in a more complex pattern which partly
involving both cortical hemispheres \cite{Carrera_2014}. In a
supplemental analysis, we showed that considering both the hemispheres
as affected significantly decreases the predictive power of the temporal triangle $\theta_T$ coefficients (\textbf{Tab S3}). These findings confirmed that the formation of clustering connections after unilateral lesions to white matter or cerebellum were actually taking place within the corresponding affected hemisphere in the cortex.

Finally, head motion and hemodynamic lags can both alter resting-state fMRI FC \cite{Power_2012,Satterthwaite_2012} and impact FC-behavior relationships \cite{Siegel_2017a}.
While both motion and hemodynamic lags are areas of ongoing methods development, the data used in this study have been carefully censored and cleaned following state-of-the art procedures, including head motion and frame-to-frame fMRI signal intensity change thresholds, tissue-based timecourse regression and excluding subjects showing severe hemodynamic disruption (hemodynamic lag $>1$ s) (more details in \cite{Siegel_2017}). 


\subsection*{Conclusion}

{\label{Discu_conclusion}}

Consistent with our hypothesis, we have identified two significant temporal network signatures that characterize dynamic brain networks after stroke. 
The formation of both clustering connections within the perilesional tissue and interhemispheric interactions with the homotopic regions are significantly abundant in stroke patients as compared to healthy controls. 
These temporal signatures, which are respectively related to intrahemispheric segregation and interhemispheric integration, varied over individuals during the subacute phases of stroke and were specific predictors of language outcome in the chronic phase. 
Furthermore, we reported a general framework for the statistical validation of hypothesized connection rules in time-varying complex networks. Taken together, our results offer new insights into the crucial role of temporal connection mechanisms in the prediction of the system performance.

\section*{Material and Methods}

{\label{MaterialMeth}}

\subsection*{Temporal exponential random graph models (tERGMs)}

To evaluate the relevance of the hypothesized local connection mechanisms in generating the observed time-varying networks, we adopted a statistical framework based on temporal exponential random graph models (tERGM) \cite{Hanneke_2010}. 

Let ${A^{0}, ... ,A^{T}}$ be a time-ordered series of graphs with $N$ fixed nodes that represents a temporally dynamic network.
By assuming one-step time dependencies, the probability of observing the entire sequence of graphs can be then written as


\begin{equation}
P(A^{1},A^2,...,A^{T} \lvert A^{0})= \prod_{t=1}^{T} P(A^{t} \lvert A^{t-1})
\label{Eq:GeranlTERGM}
\end{equation}.

The transition probability $P(A^{t} \lvert A^{t-1})$ has the following exponential form

\begin{equation}
P(A^t \lvert A^{t-1},\boldsymbol{\theta})=\frac{exp(\boldsymbol{\theta}^{'}\boldsymbol{g} (A^t,A^{t-1}))}{Z(\boldsymbol{\theta} )} 
\label{Eq:timewiseTERGM}
\end{equation}

where $\boldsymbol{\theta}$ is the vector of $r$ model parameters which weight the different graph metrics (or statistics) $\boldsymbol{g}=[g_1,g_2,...,g_r ]$, and $Z$ is a normalizing constant estimated over the space of all the graphs of size $N$. 

To incorporate temporal dependencies between time steps graph metrics can incorporate memory terms as a function of consecutive graphs.
Here, we define the temporal formation of edges between components ($E$) and the formation of triangles within modules ($T$) as

\begin{equation}
E = \sum_{ij} (1-A^{t-1}_{ij})A^{t}_{ij} (1-\delta_{ij})
\label{Eq:termEdges}
\end{equation}

\begin{equation}
T = \sum_{ijk}A^{t-1}_{ik}A^{t-1}_{kj}(1-A^{t-1}_{ij}) A^t_{ij} \delta_{ik}\delta_{kj}\delta_{ij}
\label{Eq:termTriang}
\end{equation}

where $\delta_{ij}=1$ if node $i$ and $j$ belong to the same component or module. In brain networks, components correspond to the cortical hemispheres so that $E$ quantifies the formation of interhemispheric interactions and $T$ the formation of within-hemisphere clustering connections (\textbf{Fig. 1B}). 

Because the parameter values cannot be obtained analytically, due to the computational intractability of the normalizing constant $Z$, numerical approximations are typically employed. Here, we used Markov-chain MonteCarlo maximum likelihood estimates (MCMCMLE) which is relatively fast and robust for short network sequences as compared to maximum pseudolikelihood estimates (MPLE) \cite{Leifeld_2019}.

To ensure the convergence to a meaningful solution, we introduced two secondary parameters in the tERGM, namely the instantaneous connection density $L$ and the stability $S =  \sum_{ij} A_{ij}^t A_{ij}^{t-1}+(1-A_{ij}^t)(1-A_{ij}^{t-1})$, which measures the number of persisting dyads (tied or not) between two consecutive time points. These metrics have been shown to help avoiding combinations of parameter values leading to degenerate simulations (i.e. full or empty graphs) \cite{Leifeld_2019}. 

Hence, the transition probability that we used in Eq. \ref{Eq:GeranlTERGM} to fit our data reads as

\begin{equation}
P(A^t \lvert A^{t-1},\boldsymbol{\theta})=\frac{exp(\theta_L L +\theta_E E + \theta_T T + \theta_S S}{Z(\boldsymbol{\theta})} 
\label{Eq:corrWS_TERGM}
\end{equation}

The estimated parameter coefficients can be interpreted as the (log-odds) likelihood of establishing an edge given the rest of the network and up to previous ones \cite{Leifeld_2019}. $\theta$ values can be negative or positive, with higher values indicating that the connection mechanism measured by the graph metric occurs in the network transition more often than we would expect by chance alone.

To evaluate the adequacy of the fit, we compared the actual network sequence with the ones generated by the tERMG drawing new samples from the probability function $P$. 
Here, we generated $100$ simulated network sequences.
First, we assessed the extent to which the links in the temporal network are predicted accurately by the generated simulations in each time step. We measured the prediction performance by using receiver operating characteristic and precision-recall curves and by computing the respective area under the curve AUP and AUR \cite{Leifeld_2019}. 
Then, we validated the goodness of fit by comparing the simulated values of graph quantities that were not explicitly included in the tERGM to their observed counterparts \cite{Hanneke_2010,Betzel_2017,Obando_2017}. 
Here, we used the modularity index $Q$ as defined in \cite{Newman_2006}, which intuitively captures both the integration and segregation properties of a network. 

\subsection*{Network model of brain reorganization after stroke}

We tested our tERGM on a series of time-varying networks generated by a dynamic toy-model where we could control for the formation of temporal edges and triangles over time.
The initial network $A^{0}$ consists of a graph with two disconnected modules (i.e. the hemispheres) having the same size $N$ and density $L$. This condition would represent the extreme limit for the acute effects of a stroke, i.e. low intrahemispheric segregation and low interhemispheric integration.

To simulate the reestablishment to a normal condition, we imposed the formation of inter-module edges and intra-module triangles as time elapses.
For each time-step $t=1, 2, ..., \tau$ each link of the previous network $A^{t-1}$ will be selected with a probability $q$ and rewired to form either an inter-module edge or an intra-module triangle.


Specifically, for each selected link        
 \begin{enumerate}
\item[] \textbf{if} $random(0,1) < p$
\item[(a)]  reassign it between two random nodes $(i,j)$ belonging to different modules for which $A_{ij}=0$
 \item[] \textbf{else} 
 \item[(b)] reassign it between two random nodes $(i,j)$ in the same module for which $A_{ik}=A_{kj}=1$ and $A_{ij}=0$			
 \end{enumerate} 
 
When $p=1$, only inter-module connections are formed over time, while when $p=0$ only intra-module triangles are formed and the modules are kept disconnected.
Because the links can be rewired multiple times depending on the random selection, we could generate arbitrary long network sequences.
Without loss of generality, we considered here undirected and unweighted modules with $N=50$, $L=0.2$, $\tau=5$, $q=0.2$, and $p=0.5$ as model parameters. Note that the size of the entire network is $2\times N=100$ and its density is $0.0495$ ($\sim L/2$ for large $N$), which are comparable to the values of the actual brain networks.




\subsection*{Experimental protocol and data acquisition}
The experimental data were taken after permission from a longitudinal cohort of stroke patients and healthy subjects used in a recently published study \cite{Siegel_2016,Siegel_2018}. We remind to those papers for all details related to data acquisition and processing, ethical issues, clinical and demographic information.
For the purposes of this work, we considered a group of $49$ first time human stroke patients with clinical evidence of motor, language, attention, visual, or memory deficits based on neurological examination and a group of demographically matched healthy controls ($n = 21$). 
Lesions were unilateral ($26/49$ in the left hemisphere), $23$ of them occurring at the level of the cortex (\textbf{Fig. \ref{fig3}B)} and $26$ at subcortical level including cerebellum, white matter and brainstem (\href{Table_Demo_clin.xlsx}{\textbf{File S1}}).

All subjects were recruited and underwent the same neuroimaging and behavioral exams at the Washington University School of Medicine (WUSM). Stroke patients had three longitudinal visits, i.e. 2 weeks, 3 months and 1 year after the stroke onset. Healthy controls only had two longitudinal visits at a distance of about 3 months. At each visit, resting-state functional magnetic resonance imaging (rs-fMRI) data were acquired by a Siemens 3T Tim-Trio scanner with a standard 12-channel head coil and with gradient echo EPI sequence (TR = 200 msec, TE = 2 msec, 32 contiguous 4 mm slices,$ 4 \times 4$ mm in-plane resolution). The acquisitions were six to eight resting state fMRI runs, each including 128 volumes (30 min total). 

Neuropsychological behavioral data were measured by assessing six different functional domains i.e., 1) spatial attention - assessing visual attention to the contralesional hemifield, 2) spatial memory, 3) verbal memory, 4) global language - both comprehension and production, 5) contralesional motor, and 6) contralesional visual field.
Scores in each domain were normalized to have a mean of 0 and standard deviation of 1 in controls, with lower scores indicating a greater deficit (see \textbf{Fig. \ref{fig3}D}).

\subsection*{fMRI preprocessing and functional connectivity}

For each patient, lesions were manually segmented using structural MRI images (T1-weighted MP-RAGE, T2-weighted spin echo images, and FLAIR images obtained 1-3 weeks post-stroke) using the Analyze biomedical imaging software system (\url{www.mayo.edu}).
Preprocessing of fMRI data included: 1) compensation for asynchronous slice acquisition; 2) elimination of odd/even slice intensity differences; 3) whole brain intensity normalization; 4) removal of distortion using synthetic field map estimation and spatial realignment within and across fMRI runs and 5) resampling to 3 mm cubic voxels in atlas space. Cross-modal image registration was accomplished by aligning image gradients.

Following cross-modal registration, data were passed through several additional preprocessing steps: 1) tissue-based regressors were computed based on FreeSurfer segmentation \url{surfer.nmr.mgh.harvard.edu/}; 2) removal by regression of the following sources of spurious variance; 3) temporal filtering retaining frequencies in the $.009$-$.08$ Hz range; and 4) frame censoring.
Surface generation and processing of functional data followed procedures similar to \cite{Glasser_2013}, with additional consideration for cortical segmentation in stroke patients. The left and right hemispheres were then resampled to $164000$ vertices and registered to each other \cite{VanEssen_2001}, and finally down-sampled to $10242$ vertices each for projection of functional data. fMRI signals were then smoothed using a $3$ mm Gaussian kernel.
 
The cortical surface was parcellated according to the Gordon \& Laumann atlas which includes 324 regions of interest (ROI) \cite{Gordon_2014}. 
To generate parcel-wise connectivity matrices, time-courses of all vertices within a ROI were averaged. Functional connectivity (FC) was then computed between the signals of each ROI using Fisher z-transformed Pearson correlation. All vertices that fell within the lesion were masked out, and ROIs with greater than $50\%$ lesion overlap were excluded from all analyses together with their connections. To increase the interpretation of the results, we eventually excluded ROIs that were not assigned to known systems (e.g., visual, salience, default, etc). 
As a result, we obtained FC connectivity matrices with slightly different number of nodes ($\left\langle N \right\rangle=275.6$, sd=$6.8$).

\subsection*{Brain network construction and modeling}
ERGMs assume that there is a homogeneous process operating on the entire network. In stroke, the reconfiguration processes are mainly taking place within and between the affected regions of the brain. To take into account such heterogeneity and improve the specificity our approach we modeled for each patient only the network impacted by the damage. 
For cortical lesions, we considered only the connections among the ROIs of the systems directly affected by the lesion (e.g. \textbf{Fig. \ref{fig3}B)}. This procedure gave in average connectivity matrices with smaller size ($\left\langle N \right\rangle=81.13$, sd$=47.78$, \href{Table_Demo_clin.xlsx}{\textbf{File S1}}).
For subcortical lesions - including damages to cerebellum, white matter, and brainstem - we considered that all the ROIs were potentially affected and we preserved the entire connectivity matrix. To ensure a fair comparison we also considered the whole connectivity matrix for the healthy controls.

Because tERMGs have been mainly studied for unweighted networks, we thresholded the connectivity matrices to retain a same percentage of strongest links in each brain network. While loosing information, this procedure has the advantage to ensure a more robust comparison across different subjects and conditions \cite{De_Vico_Fallani_2014}. Specifically, we considered a connection density of $0.1$, which falls in the range of typically studied values, i.e. ($0.05$, $0.2$) \cite{Siegel_2018,Lord_2012,DeVicoFallani_2017} .
The resulting sparse time-varying brain networks were represented by adjacency matrices $A^{t}$, where each entry indicated the presence $A^{t}_{ij}=1$ or the absence $A^{t}_{ij}=0$ of a link between nodes $i$ and $j$ at the visit $t$. 

We modeled every network sequence through a tERGM as described in Eq. \ref{Eq:corrWS_TERGM}. Because the graph metric $T$ counts the number of triangles formation within the affected hemisphere, we specified this information in the model by restricting opportunely the sum indices in Eq. \ref{Eq:termTriang}. For cortical lesions the affected hemisphere corresponded to the lesion side. For subcortical-lesioned patients, the affected hemisphere corresponded to the lesion side if the damage was in white matter, while we considered the contralesional hemisphere if the stroke was in the cerebellum; if the damage was in the brainstem we labelled both the cortical hemispheres as affected.
Finally, because we were interested in the strict formation of triangles over time we added the product term $(1-A^{t}_{ik})(1-A^{t}_{kj})$ in Eq. \ref{Eq:termTriang}. This term imposes that complete triangles can never occur in one step; instead the closure of triangles can only exist longitudinally.

\section*{Acknowledgments}
Authors would like to acknowledge Gordon Shulman for the design/collection of the experiemntal data used in this work.
The research leading to these results has received funding from the
program ``Investissements d'avenir'' ANR-10-IAIHU-06 (Agence Nationale
de la Recherche-10-IA Institut Hospitalo-Universitaire-6). FD
acknowledges support from the ``Agence Nationale de la Recherche''
through contract number ANR-15-NEUC-0006-02. MC acknowledges support from the ``National Institute of Health" through contracts number R01-HD06117 and R01-NS095741.
The content is solely the responsibility of the authors and does not necessarily represent the
official views of any of the funding agencies.

\nolinenumbers

\par\null

\clearpage
\bibliographystyle{plos2015}
\nolinenumbers
\begin{small}
\bibliography{converted_to_latex.bib}
\end{small}

\end{document}